\documentclass[twocolumn,preprintnumbers,amsmath,amssymb,pra,showkeys]{revtex4}

\usepackage{graphicx}
\usepackage{dcolumn}
\usepackage{bm}

\usepackage{float}
\usepackage{hyperref}
\usepackage{mathrsfs}
\usepackage{color}
\usepackage{pgfplots}
\usepackage{filecontents}

\begin{document}


\title{Spontaneous Emission in Quantum Walks of a Kicked Bose-Einstein Condensate}

\author{Caspar Groiseau$^{1,4}$}
\author{Sandro Wimberger$^{1,2,3}$}%
 \email{sandromarcel.wimberger@unipr.it}
\affiliation{%
$^{1}$ITP, Heidelberg University, Philosophenweg 19, 69120 Heidelberg, Germany.\\
$^{2}$Dipartimento di Scienze Matematiche, Fisiche ed Informatiche, Universit\`a di Parma, Parco Area delle Scienze 7/A, 43124 Parma, Italy\\
$^{3}$INFN, Sezione di Milano Bicocca, Gruppo Collegato di Parma, Parma, Italy\\
$^{4}$Dodd-Walls Centre for Photonic and Quantum Technologies, Department of Physics,\\
 University of Auckland, Private Bag 92019, Auckland, New Zealand
}%


\date{\today}

\begin{abstract}
We analytically investigate the recently proposed and implemented discrete-time quantum walk based on a kicked Bose-Einstein condensate. We extend previous work on the effective dynamics by taking into account spontaneous emission due to the kicking light. Spontaneous emission affects both the internal and external degrees of freedom, arising from the entanglement between them during the walk dynamics. The result is a measurable degrading of the experimental walk signal that we characterise.
\end{abstract}

\maketitle

\section{Introduction}
\label{intro}

Quantum walks \cite{PhysRevA.48.1687,doi:10.1080/00107151031000110776,Buch2013} vastly differ from their classical analogue of random walks by exhibiting interference effects in their probability distribution. The possible practical applications of quantum walks in the fields of quantum information \cite{DT-QC} and quantum metrology \cite{Buch2013} have lead to a surge in interest in the development of a series of proposed schemes and experimental implementations \cite{Karski174, PhysRevA.96.023620, PhysRevLett.104.050502, PhysRevLett.100.170506, PhysRevLett.103.090504, PhysRevLett.108.010502, Cardano2017}. In particular, discrete-time quantum walks with cold atoms in optical lattices were proposed in \cite{PhysRevA.66.052319} realized later in \cite{Karski174}. 

We are interested in the recent idea of using the atom optics kicked rotor to perform walks in momentum space \cite{Weiss2015, PhysRevA.93.023638}. A working experiment with a Bose-Einstein condensate is currently being run at Oklahoma State University \cite{PhysRevLett.121.070402}. This realization is quite stable and allows for the implementation of a few tens of walk steps. Evolving longer in time, it however suffers from spontaneous emission (SE) induced by the same laser light that creates the kicking potential. Here, we introduce a realistic model to include the affects of SE on the walk dynamics and on its coherence. \textcolor{black}{Our model substantially extends previous work on the atom-optics kicked rotor (AOKR) for which two levels suffice.} It contains three internal states, one excited state and the two ground hyperfine levels used for the internal degree of freedom that determines the walker's direction. An event of SE induces a photon recoil in the external, momentum degree of freedom of the atoms, thus changing their (quasi)momentum. This affects the walk realization which is based on specific values of quasimomentum to engineer the directed ratchet-like motion \cite{PhysRevA.93.023638}. A secondary effect is the collapse of the wave function after SE, projecting the superposition of internal states \textcolor{black}{onto a specific one of the two} hyperfine level.

The paper is organised as follows: the next sec. \ref{sec:2} reviews the closed system realising a discrete-time quantum walk of kicked atoms. Sec. \ref{sec:3} presents our theoretical model for the three-state system and its complete temporal evolution. Numerical results based on our model are shown in sec. \ref{sec:4}. The conclusions are finally found in sec. \ref{concl}.

\section{Review of the closed system's evolution}
\label{sec:2}

\begin{figure}[tb]
	\includegraphics[width=\linewidth]{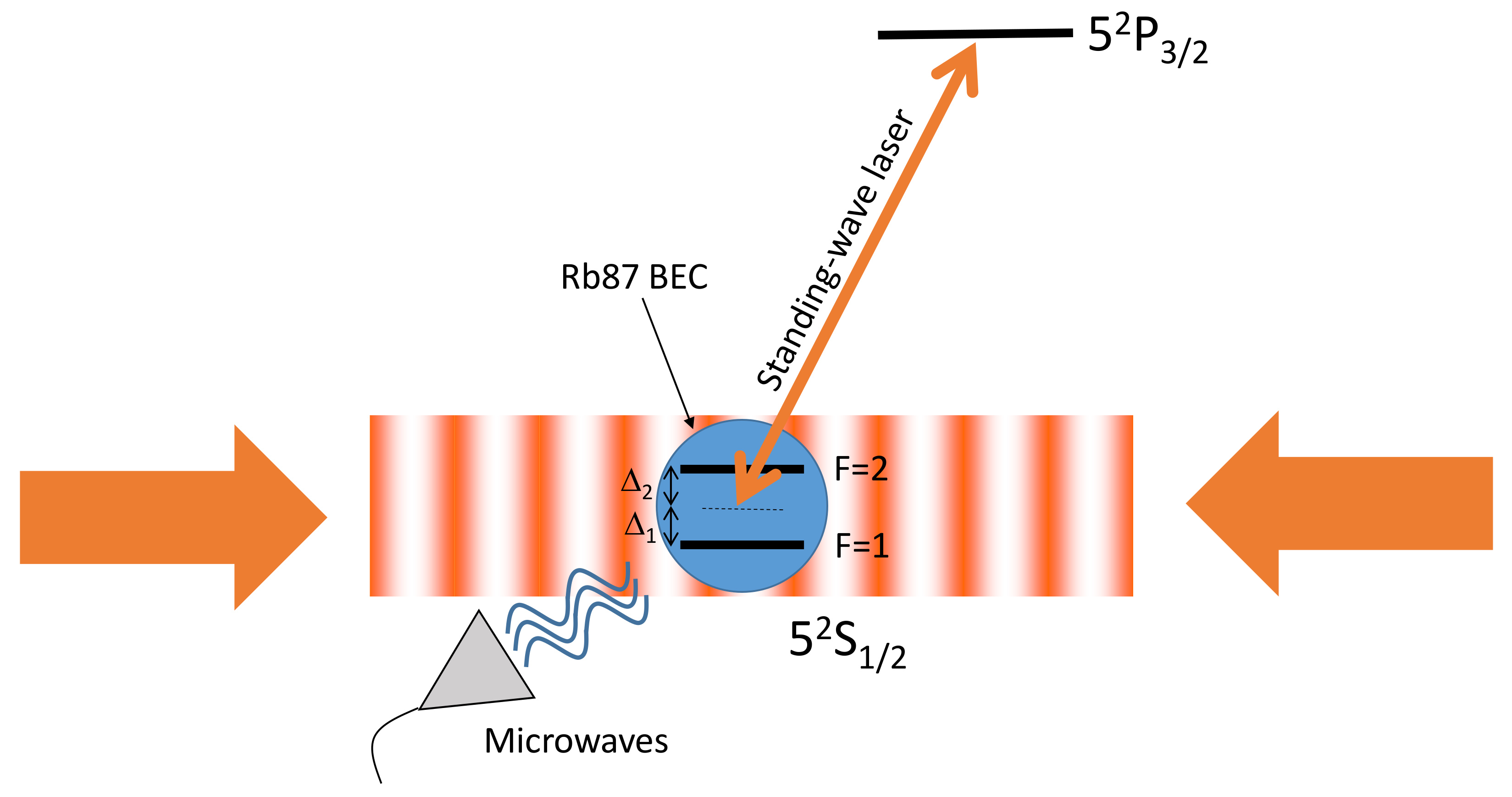}
	\caption{Schematic representation of the experimental implementation, adapted from \cite{PhysRevA.93.023638}. 
	The optical lattice is pulsed periodically to implement the momentum shifts at quantum resonance conditions. The internal grounds states $F = 1$ and $F = 2$, together with the excited $P_{3/2}$ state are necessary to control the quantum walk. They also constitute the three-levels of the $^{87}$Rb atoms in our model for SE. The detunings $\Delta_1$ and $\Delta_2$ our usually chosen equal for a symmetric quantum walk.}
	\label{fig1}
\end{figure}

The experimental quantum walk is implemented with a Bose-Einstein condensate of ultra-cold $^{87}$Rb atoms. \textcolor{black}{In contrast to the standard AOKR, which effectively works with only one internal level \cite{Nonl2003, SW2011},} here two ground state hyperfine levels $F=2$ (denoted by $|2\rangle$ in the following) and $F=1$ (denoted by $|1\rangle$) form the internal degree of freedom that define the 'coin' space. The external degree of freedom is the centre-of-mass momentum of the atoms. The atoms are periodically kicked by a standing-wave laser of frequency $\omega$ and period $\tau$. The laser is tuned from the excited state manifold $|e\rangle$ between the two ground states (see Fig. \ref{fig1} for a schematic representation).

During the kick the system may be described by the following Hamiltonian in rotating wave approximation and in the time-independent frame
\begin{align}
	H&=\frac{\hbar\Omega}{2}\cos\frac{\theta}{2}\left(|1\rangle\langle e|+|e\rangle\langle 1|\right) \\
	&+\frac{\hbar\Omega}{2}\cos\frac{\theta}{2}\left(|2\rangle\langle e|+|e\rangle\langle 2|\right).
\end{align}
Note that we removed the ground state dynamics $H_g=\sum_{m=1}^{2}E_m|m\rangle\langle m|,E_1=-\hbar\Delta_1,E_2=\hbar\Delta_2$ (likewise in the rest of this paper), as the resulting dynamical phase shift is assumed to be compensated at all times. $\Delta_{1,2}$ are the detunings of the levels from the resonant excitation transition in Fig. \ref{fig1}. We do not put heads on top of operators, but since we work with a completely quantum mechanical problem the Hamiltonian, the evolution and angle and momentum variables are to be read as operators.

In earlier work \cite{2018arXiv180205476G}, we derived the effective dynamics during the kick by adiabatically eliminating the excited state using the method of James and Jerke \cite{doi:10.1139/p07-060} leading to the effective Hamiltonian
\begin{equation}
\label{Heff}
H_{\rm eff}=-\frac{\hbar\Omega^2}{4\Delta_1}\cos^2\frac{\theta}{2}|1\rangle\langle 1|+\frac{\hbar\Omega^2}{4\Delta_2}\cos^2\frac{\theta}{2}|2\rangle\langle 2| \,.
\end{equation}
As detailed in ref. \cite{2018arXiv180205476G}, this leads to a conditional kick operator in the space spanned by the two internal states $|1\rangle$ and $|2\rangle$
\begin{equation}
\label{kickmatrix}
K=\left(\begin{array}{cc}
e^{ik_1\cos\theta} & 0 \\
0 & e^{-ik_2\cos\theta} \\
\end{array}\right).
\end{equation}
where the kick strengths $k_m$ are given by
\begin{equation}
\label{kickstrength}
k_m=\frac{\Omega^2\tau_p}{8\Delta_m} \,.
\end{equation}
Here, we used the Rabi frequency $\Omega$, the finite duration of the kick pulse $\tau_p$ and the detuning of the laser $\Delta_m$.

As usual for the kicked-rotor model, the kick is assumed to be so short that the free phase evolution given by
\begin{equation}
F=e^{-i\tau\frac{ p^2}{2}}\,,
\end{equation}
can be neglected during the kick , i.e., $\tau_p \ll \tau$. In our dimensionless units, $p=n+\beta$, with $n \in \mathbb{Z}$ and $\beta \in [0,1)$ being the conserved quasimomentum in the periodic kicking potential \cite{Nonl2003, SW2011}.

We consider the system to be in quantum resonance \cite{1990PhR...196..299I} so that the kick period $\tau$ is chosen in such a way that the free evolution completely rephases in momentum space from one kick to the next one. The latter rephasing condition depends also on the quasimomentum of the atom \cite{Nonl2003, SW2011}. For perfect 'resonant' choice of quasimomentum, the free evolution can be neglected even in-between two kicks. Slight deviations from the conditions for resonant quasimomenta induce a systematic degradation of the ideal quantum resonant motion \cite{Wimberger2005}, and hence of the walk evolution as well.

The initial state of the atoms in momentum space is a quantum ratchet configuration \cite{sadgrove2007, gil2008, PhysRevA.94.043620,doi:10.1002/andp.201600335}, a superposition of integer momentum classes like
 \begin{equation}
 	|\psi\rangle=\frac{1}{\sqrt{2}}(|n=0\rangle-i|n=1\rangle)
 	\label{initial_state}
 \end{equation}
 for which the momentum distribution propagates asymmetrically in momentum space rather than diffusing symmetrically around the initial state like a single momentum class would. The direction of the propagation depends on the sign of the kick strength, which itself is inversely proportional to the detuning of the laser \cite{sadgrove2007, SW2011}.
 
The internal states are addressed by the two-parameter unitary rotation matrix, which in the experiment is controlled by microwaves. We start by creating an equal superposition of both hyperfine states
 \begin{equation}
 |\Psi\rangle=\frac{1}{\sqrt{2}}\left(|1\rangle+|2\rangle\right)\otimes|\psi\rangle.
 \end{equation}
 Then after each of the kicks, see Eq. \eqref{kickmatrix}, we mix these internal levels by applying the following 50:50 beam splitter coin toss
 \begin{equation}
 \label{mix}
 C=\frac{1}{\sqrt{2}}\left(\begin{array}{cc}
 1 & i \\
 i & 1 \\
 \end{array}\right).
 \end{equation}
 
The main experimental observable is the total momentum distribution. It is computed from the sum of the momentum distribution of the two ground states
\begin{equation}
	P(n;T)=P_1(n;T)+P_2(n;T)
	\label{totaldistribution}
\end{equation}
For more details on the realization of the system and experimental results we refer to \cite{PhysRevLett.121.070402}.

\section{Spontaneous Emission}
\label{sec:3}

The experiment, not being a perfectly closed system, suffers from loss of coherence through diverse decoherence channels. Spontaneous emission (SE), the random relaxation of an excited atom is one of the most important ones of these effects, at least on longer timescales. It becomes more important the bigger the kick strength $k$ or the smaller the detuning $\Delta_m$ ($m=1, 2$) is. 

SE led to many measurable effects in AOKR systems \cite{PhysRevLett.81.1203, Ammann1998, 1464-4266-1-4-301, 1464-4266-2-5-307, daleySpontem, darcy2001, PhysRevE.64.056233, PhysRevE.66.056210, DarcyPRE, Nonl2003, PhysRevE.88.034901}. All the latter realisations were not really sensitive to the specific internal ground state and could be well modelled by taking into account just one of them. In our quantum walk, the situation is different since the two different hyperfine levels determine the different directions of the walker. \textcolor{black}{Then SE crucially affects the time evolution of the walk, and, in contrast to the standard AOKR \cite{Nonl2003}, analytical solutions like in the coherent closed case \cite{2018arXiv180205476G} seem impossible. Even the numerical evolution becomes more involved due to the more complex Lindblad operators to be introduced in the next section.}

\subsection{The model for spontaneous emission}
\label{sec:3_1}

Our model for SE is based on the three states shown in Fig. \ref{fig1}: one excited state and the two hyperfine ground states. Loss to other external channels will be neglected here. Below it will become clear that a SE event will shift also the quasimomentum away from its optimal resonant value for our walk. This implies that the shifted atoms effectively won't take part any more in the directed walk evolution. As this effect essentially models a loss channel for the walker, our assumption of a closed three level system seems justified.  

The dynamics may be described by a Lindbladian dissipator acting on the atomic density operator
\begin{equation}
	\label{ME}
	\mathcal{D}[\rho]=\sum_{m=1}^{2}-\frac{1}{2}\left(L_m^\dagger L_m\rho+\rho L_m^\dagger L_m-2L_m\rho L_m^\dagger\right),
\end{equation}
with the Lindblad operators
\begin{equation}
	L_m=\sqrt{\gamma_m}|m\rangle\langle e|.
\end{equation}
The $\gamma_m$ are the spontaneous emission rates \cite{PhysRevE.88.034901} and can be computed from
\begin{equation}
	\gamma_m=\frac{k_m}{\tau_p\tau_{SE}\Delta_m},
\end{equation}
where $\tau_p$ is the pulse duration and $\tau_{SE}$ the lifetime of the transition. We also define the total decay rate as the sum of these two decay rates
\begin{equation}
	\gamma=\gamma_1+\gamma_2
\end{equation}
 and the corresponding probability of spontaneous emission per kick
 \begin{equation}
 	p_{\rm SE} = \gamma\tau_p.
 \end{equation}

\subsection{The Recoil motion of the spontaneously emitting atom}
\label{sec:3_2}

Up to now we have ignored the effect of spontaneous emission on the motional state of the atom. During the event, the atom collapses from the excited state $|e\rangle$ onto one of the ground states $|1\rangle$ or $|2\rangle$ by emitting a photon of the corresponding energy difference. The energy difference is different for both decoherence channels and so the different wave vectors $\kappa_m$ will to lead to two distinct recoil momenta
\begin{align}
	\hbar\kappa_1&=\hbar\frac{\omega+\Delta_1}{c}\\
	\hbar\kappa_2&=\hbar\frac{\omega-\Delta_2}{c}.
\end{align}
Since the standing-wave laser frequency exceeds its detuning $\omega\gg\Delta_m$ by orders of magnitude, we may set all the shifts approximately equal
\begin{equation}
\kappa_1=\kappa_2=\kappa=\frac{\omega}{c}.
\end{equation}
For a standing-wave laser polarized in $z$-direction, the direction of the momentum shift is randomly distributed according to \cite{PhysRevA.53.2522}
\begin{equation}
	\Xi(\phi,\theta)=\Xi(\theta)=\frac{3}{8\pi}\left[1-\cos^2\theta\right], \phi\in[0,2\pi], \theta\in[0,\pi].
\end{equation} 
Since only the projection $u$ along the walk axis ($x$-axis) matters to us we compute its distribution which amounts to
\begin{equation}
\Xi(u)=\frac{3}{8}\left[1+u^2\right], u\in[-1,1].
\end{equation}
In the end, we add a recoil term to each Lindblad operator of the master equation
\begin{equation}
\label{addition}
L_m\rightarrow L_me^{-iu \frac{\theta}{2}}\,.
\end{equation}
The shift by $u$ affects quasimomentum $\beta$, and possibly also the integer parts of momentum $n$, as largely discussed in \cite{Nonl2003}. The corresponding terms in the master equation must be integrated over the $u$-component of the recoil momentum.


%

\section{Effective Dynamics during the Kick}
\label{sec:4}

\subsection{Effective Linblad operator formalism}
\label{sec:4_1}

We are now interested in eliminating the excited state in the decoherent part of our evolution. For this purpose, we use the effective Lindblad operator technique of Reiter and S\o{}rensen \cite{PhysRevA.85.032111}, which is a combination of Feshbach projection operator formalism \cite{FESHBACH1958357} and perturbation theory.

The prerequisites for applying this effective Lindblad operator formalism are:
\begin{itemize}
	\item There are no initial excitations to the upper level in the system. Excitations only get introduced into the system by the kick. The lifetime of the excited state is much smaller than the finite kick pulse length and even more as compared with the kick period ($\tau_{SE}\ll\tau_p \ll\tau$). Then the excitations from an earlier kick should easily have relaxed by the start of an upcoming kick. Hence, this approximation is more than justified.
	\item To be able to perform perturbation theory the interaction between the excited and ground states has to be sufficiently weak. The Rabi frequency $\Omega$ represents the strength of our coupling, for a typical kicked-rotor experiment it is in the order of magnitude of $1$ GHz, which is sufficiently lower than the atomic transition frequencies that lie in the optical regime \cite{Steck}.
	\item The lifetime of the excited state has to be short enough, so that we can perform adiabatic elimination, i.e., we assume that on average the excited state is not populated. With a lifetime of $26$ ns \cite{Steck} this should be guaranteed.
\end{itemize}
We introduce the projectors onto the excited and ground states
\begin{align}
	P_e&=|e\rangle\langle e|\\
	P_g&=|1\rangle\langle 1|+|2\rangle\langle 2|.
\end{align}

With these projectors the Hamiltonian is separated into four parts, two describing the ground (same as $H_g$, disregarded here) and excited states and two describing transitions between these two
\begin{align}
	H_e&=P_eHP_e=0\\
	V_-&=P_gHP_e=\underbrace{\frac{\hbar\Omega}{2}\cos\frac{\theta}{2}|1\rangle\langle e|}_{V_-^{(1)}}+\underbrace{\frac{\hbar\Omega}{2}\cos\frac{\theta}{2}|2\rangle\langle e|}_{V_-^{(2)}}\\
	V_+&=P_eHP_g=\underbrace{\frac{\hbar\Omega}{2}\cos\frac{\theta}{2}|e\rangle\langle 1|}_{V_+^{(1)}}+\underbrace{\frac{\hbar\Omega}{2}\cos\frac{\theta}{2}|e\rangle\langle 2|}_{V_+^{(2)}}.
\end{align}
In addition they define a non-hermitian Hamiltonian, similar to the one encountered in the quantum jump picture \cite{Molmer:93}, that coalesces the fast-oscillating excited state and decoherent dynamics
\begin{equation}
	\mathcal{H}=H_e-\frac{i\hbar}{2}\sum_{m=1}^{2}L^\dagger_mL_m=-i\frac{\hbar\gamma}{2}|e\rangle\langle e|.
\end{equation}
\begin{widetext}

\begin{figure}[tbh]
		\centering
		\includegraphics[width=\textwidth]{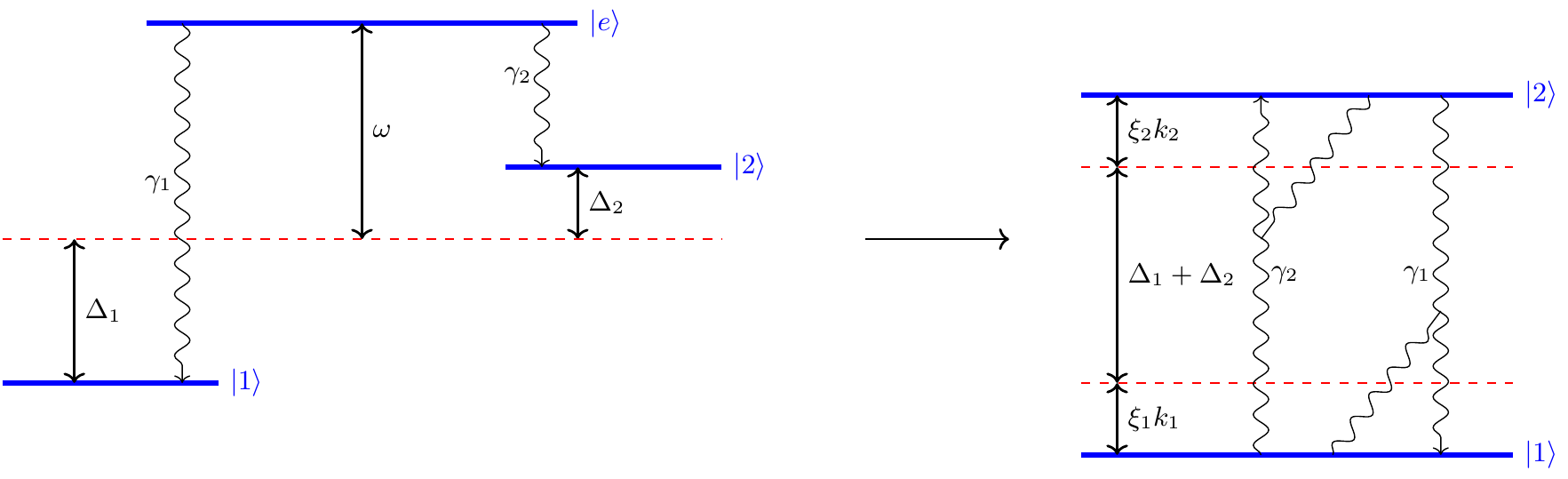}		
		\caption{Schematic representation of the system during the kick as an atomic three-level system in $\Lambda$-configuration (left) and of the effective system as an atomic two-level system (right). The adiabatic elimination of the excited state creates an additional light-shift energy difference between the levels \cite{2018arXiv180205476G} marked with the different kicking strengths $\xi_1k_1$ and $\xi_2k_2$.}
		\label{fig2}
	\end{figure}
\end{widetext}

The formalism continues with transforming the following evolution operator 
\begin{equation}
U(t)=e^{-i(H_g+\mathcal{H})t}
\end{equation}
into the interaction picture. Then, we perform perturbation theory: the density operator is expanded in terms of the interaction strength $\epsilon\propto \tilde V_+ + \tilde V_-$, where the operators with tildes are understood in the interaction picture:
\begin{equation}
\tilde\rho=\tilde\rho^{(0)}+\epsilon\tilde\rho^{(1)}+\epsilon^2\tilde\rho^{(2)}+...
\end{equation} 
Without initial excitations adiabatically eliminating the excited state consists in setting
\begin{equation}
	P_e\tilde\rho^{(2)}P_e\approx 0.
\end{equation}
In the end, we get an effective master equation no longer involving the excited state. From this equation one can read off the new effective Hamiltonian and Lindblad operators. The effective Hamiltonian is then
\begin{equation}
\begin{split}
\label{neweff}
H_{\text{eff}}&=-\frac{1}{2}\bigg[V_-\sum_{m}\frac{1}{\mathcal{H}-E_{m}}V_+^{(m)}\\
&+\sum_{m}V_-^{(m)}\frac{1}{\mathcal{H}^\dagger-E_{m}}V_+\bigg]\\
&=-\frac{\Delta_1\Omega^2\cos^2\frac{\theta}{2}}{4\Delta_1^2+\gamma^2}|1\rangle\langle 1|+\frac{\Delta_2\Omega^2\cos^2\frac{\theta}{2}}{4\Delta_2^2+\gamma^2}|2\rangle\langle 2|\\
&+\frac{(\Delta_2-\Delta_1)\Omega^2\cos^2\frac{\theta}{2}}{8(\Delta_1+i\frac{\gamma}{2})(-\Delta_2-i\frac{\gamma}{2})}|1\rangle\langle 2|\\
&+\frac{(\Delta_2-\Delta_1)\Omega^2\cos^2\frac{\theta}{2}}{8(\Delta_1-i\frac{\gamma}{2})(-\Delta_2+i\frac{\gamma}{2})}|2\rangle\langle 1|.
\end{split}
\end{equation}
Here, the last two terms represent an effective coupling between the ground states that can be ignored as they scale with the difference of the two detunings $(\Delta_2-\Delta_1)$ which is small. Indeed, most of the experimental data was taken for equal detunings $\Delta_2=\Delta_1$ that produces a perfectly symmetric walk \cite{PhysRevLett.121.070402}. Biased walks are also possible with slightly different detunings, and hence different kicking strengths \cite{PhysRevA.93.023638, PhysRevLett.121.070402}. 

Altogether, we have now
\begin{equation}
H_{\rm eff}=-\xi_1\frac{\hbar\Omega^2}{4\Delta_1}\cos^2\frac{\theta}{2}|1\rangle\langle 1|+\xi_2\frac{\hbar\Omega^2}{4\Delta_2}\cos^2\frac{\theta}{2}|2\rangle\langle 2| \,,
\end{equation}
where we defined
\begin{equation}
\xi_m=\frac{1}{1+\frac{\gamma^2}{4\Delta_m^2}}\approx 1 \,.
\end{equation}
 In the limit of no SE  $\gamma\rightarrow0$, we recover the previous Hamiltonian from Eq. \eqref{Heff}, so these two results are in agreement.

Note that the prefactor $\xi_m$ will affect the kicking strength $k_m$, and therefore also the relative light-shift phase between the two hyperfine states, see sec. \ref{sec:2} and ref. \cite{2018arXiv180205476G} for details. This new phase has to be included into the phase correction now
\begin{equation}
	\Phi_{\rm dyn}(\Delta_1,\Delta_2)=\xi_1k_1+\xi_2k_2+(\Delta_1+\Delta_2)\tau.
\end{equation}

The resulting effective Lindblad operators
\begin{align}
\mathscr{L}_m&=L_m\left(\frac{1}{\mathcal{H}-E_{1}}V_+^{(1)}+\frac{1}{\mathcal{H}-E_{2}}V_+^{(2)}\right)\\
	\begin{split}
		\label{leff1}
	&=\sqrt{\gamma_m}\cos\frac{\theta}{2}\bigg[\frac{\Omega}{2(\Delta_1-\frac{i\gamma}{2})}|m\rangle\langle 1|\\
	&+\frac{\Omega}{2(-\Delta_2-\frac{i\gamma}{2})}|m\rangle\langle 2|\bigg]	
	\end{split}	
\end{align}
have now become bipartite. They no longer solely describe the decay from the excited state to one of the ground states but rather include also the preceding excitation process. \textcolor{black}{Once a SE event has happened during a kick, the walker is no longer in a superposition of the two internal levels but is projected onto one of them. It will not return to the other level for the remainder of this kick or step of the walk (with the assumption of an instantaneous coin toss). This means that the walker then acts as a classical ratchet for this kick and propagates only into one direction until the next coin toss again mixes both internal levels. Though the direction of the motion may change if multiple SE events happen during one kick, we must average over a sufficiently large sample of numerical trajectories of the walk subject to SE in order to assure an equal mean number of SE decays onto both levels.}

The cosinusoidal position dependence in the Lindbladian is easily explained as a position dependent variation of the excitation probability, which is maximal at the potential maxima of the kicking light \cite{1464-4266-2-5-307}. This position dependence can be interpreted as a coherent momentum shift of one photonic recoil (half a natural momentum unit) along the standing-wave laser axis arising from the absorption of one photon in the excitation. 


\subsection{Results}
\label{sec:4_2}

Let us now write down the full master equation of the system. It is helpful to separate the internal and external part of the Lindbladians for comparison with master equations of kicked-rotor systems without internal dynamics. We therefore write
\begin{equation}
	\begin{split}
		\label{leff}
		\mathscr{L}_{m}&=\sqrt{\gamma_m}\mathscr{L}_{m,\text{ext}}\mathscr{L}_{m,\text{int}}\\
		\mathscr{L}_{m,\text{ext}}&=e^{-iu\frac{\theta}{2}}\cos\frac{\theta}{2}\\
		\mathscr{L}_{m,\text{int}}&=\frac{\Omega}{2(\Delta_1-\frac{i\gamma}{2})}|m\rangle\langle 1|+\frac{\Omega}{2(-\Delta_2-\frac{i\gamma}{2})}|m\rangle\langle 2|,
	\end{split}
\end{equation}
and
\begin{widetext}
\begin{equation}
\begin{split}
\label{master}
\dot\rho&=-i\left[H_{\text{eff}},\rho\right]\\
&-\frac{\gamma_1}{2}\Bigg(\mathscr{L}_{1,\text{int}}^\dagger \mathscr{L}_{1,\text{int}}\cos^2\frac{\theta}{2}\rho+\rho\cos^2\frac{\theta}{2}\mathscr{L}_{1,\text{int}}^\dagger \mathscr{L}_{1,\text{int}}-2\int_{-1}^{1} du \Xi(u)\mathscr{L}_{1,\text{int}} e^{-iu \frac{\theta}{2}}\cos\frac{\theta}{2}\rho\cos\frac{\theta}{2}e^{iu \frac{\theta}{2}}\mathscr{L}_{1,\text{int}}^\dagger\Bigg)\\
&-\frac{\gamma_2}{2}\Bigg(\mathscr{L}_{2,\text{int}}^\dagger \mathscr{L}_{2,\text{int}}\cos^2\frac{\theta}{2}\rho+\rho\cos^2\frac{\theta}{2}\mathscr{L}_{2,\text{int}}^\dagger \mathscr{L}_{2,\text{int}}-2\int_{-1}^{1} du \Xi(u) \mathscr{L}_{2,\text{int}} e^{-iu \frac{\theta}{2}}\cos\frac{\theta}{2}\rho\cos\frac{\theta}{2}e^{iu \frac{\theta}{2}}\mathscr{L}_{2,\text{int}}^\dagger\Bigg).
\end{split}
\end{equation}
\end{widetext}
This result is similar to those derived for the kicked rotor without an internal level structure \cite{1464-4266-2-5-307,PhysRevA.53.2522}. The main difference arises from the additional ground state\textcolor{black}{. Eqs. \eqref{leff}, \eqref{leff1} and \eqref{master} contain an additional dissipator.  Each of them projects onto a different internal level. The more complex dissipator, in principle, allows for more control, since the rates can be controlled to some extent by the detunings in the experiment as well as by additional offsets of the optical potential. However, as will be shown below, the ratio of the two independent SE rates has little impact on the global evolution of the measured momentum distributions. The reason is that the dominant effect on the distributions in the master equation will be the loss of the resonance condition of the individual walker introduced by the random shift of the quasimomentum.} 

In practice, we solve the master equation numerically using a Monte-Carlo wave function technique \cite{Molmer:93}. The code for the full walk is based on the quantum Floquet map of the kicked-rotor evolution \cite{wimberger2014nonlinear}, where free evolution and kick factorize due to the $\delta$-kick assumption, which are evaluated in position and momentum space, respectively. We switch back and forward these two with a Fast-Fourier-Transform (FFT) \cite{Press02a}. 

\begin{figure}[tb]
	\includegraphics[width=0.49\linewidth]{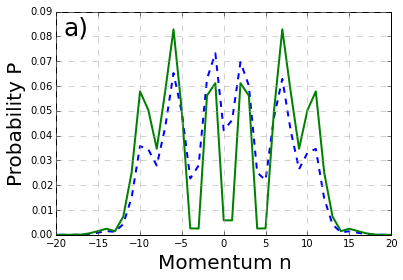}
	\includegraphics[width=0.49\linewidth]{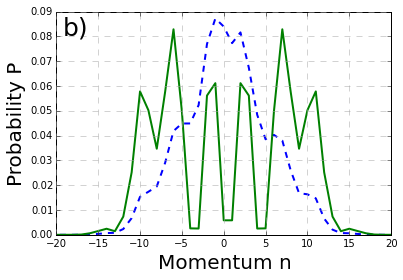}\\
	\includegraphics[width=0.49\linewidth]{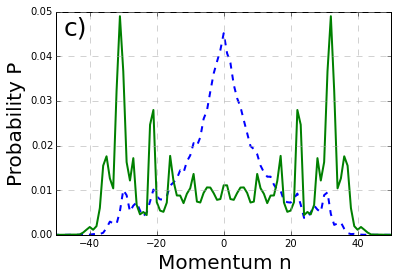}
	\includegraphics[width=0.49\linewidth]{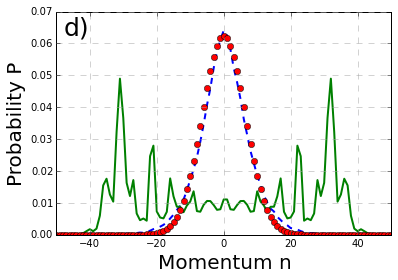}
	\caption{Numerical simulations of quantum walks for fixed resonant quasimomentum with $k=1.45$, $T=15$ (a,b) or $T=50$ (c,d) steps. The green solid lines show a walk without SE as compared with the second one including it (blue dashed line) with rates per kick $p_{\rm SE}=0.037$ (a,c) and $0.11$ (b,d), respectively. Panel (d) includes a Gaussian (red-dotted line) for the comparison with the classical limit of the walk. The walks with decoherence by SE are averages over $1000$ Monte-Carlo trajectories and have up to $60000$ split steps within the duration of one kick.}
	\label{fig3}
\end{figure}

Now since the kick  realistically has a finite length in time we have to split the kick into small steps and apply a FFT switching here as well. During each kick, we draw up to three exponentially (Poisson) distributed times at which a spontaneous emission event is deemed to happen. Kicks with more than three events are very unlikely for typical parameters and therefore statistically irrelevant \cite{Nonl2003}. If the drawn times lie inside the finite pulse width of the kick we keep them, otherwise they are discarded. At said events we apply the collapse operator from Eq. \eqref{leff}. Within the kick duration $\tau_p = 380$ ns, which is the experimental value of the recent experiment \cite{PhysRevLett.121.070402}, we apply up to $60000$ split-operator steps for the data seen in Fig. \ref{fig3}. The chosen initial state was a ratchet state from Eq. \eqref{initial_state} with one fixed resonant quasimomentum only. 

After the associated momentum shift we might need to shift the momentum so that the quasimomentum $\beta$ stays in the first Brillouin zone. This has in theory also to be done for the cosinusoidal part of the collapse operator that is nothing but a superposition of two momentum shifts of $\pm\frac{1}{2}$ in natural units. To save computation power one may replace this term by its mean value
\begin{equation}
\cos\frac{\theta}{2}=\frac{e^{i\frac{\theta}{2}}+e^{-i\frac{\theta}{2}}}{2}\approx\langle\cos\frac{\theta}{2}\rangle=\frac{1}{\sqrt{2}}.
\end{equation}

From Fig. \ref{fig3}, we see that the stronger the SE rate becomes the more the walk warps from its bimodal nature to the unimodal form of a classical random walk. The same transition happens with increasing number of steps $T$ of the walk for a fixed given SE rate.



From Figs. \ref{fig3} (a,c) we see that the limit of SE rate is about $p_{\rm SE} \le 0.05$ for up to $15\ldots20$ steps of the walk. For larger rates, the ballistic peaks are turned quickly into a distribution centered around zero, as seen in Figs. \ref{fig3} (b,c,d).

\begin{figure}[tb]
	\includegraphics[width=0.49\linewidth]{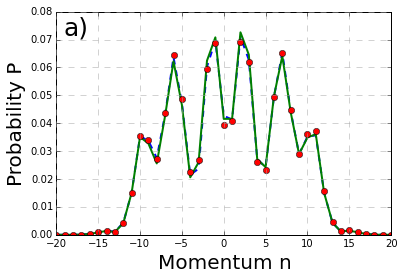}
	\includegraphics[width=0.49\linewidth]{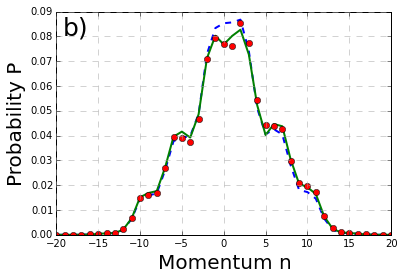}\\
	\includegraphics[width=0.49\linewidth]{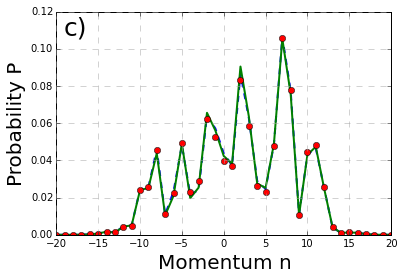}
	\includegraphics[width=0.49\linewidth]{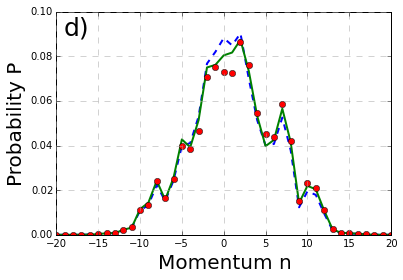}
	\caption{Numerical simulations of quantum walks for fixed initially resonant quasimomentum with $k=1.45$ and $T=15$ steps. The blue dashed lines show a walk with SE with a 50:50 chance to jump onto each level, the green solid line represents a ratio of 70:30 and the red dotted line of 99:1. The total rates per kick are $p_{\rm SE}=0.037$ (a,c) and $0.11$ (b,d), respectively. Panels (c) and (d) show the partial momentum distributions for only one of the two internal levels, whilst (a) an (b) show the usual total distributions as plotted in Figs. \ref{fig3} and \ref{fig5}.}
	\label{fig4}
\end{figure}

\textcolor{black}{The walk is very stable with respect to biased SE rates as anticipated earlier. In Fig. \ref{fig4} we demonstrate this stability of the walk under variation of the relative weight of the decay channels. Both the total and the partial momentum distribution show only minor deviations towards the origin of the distribution, which anyhow would be hard to resolve experimentally. This is in accord with our earlier statement that the dominating effect is the loss of the resonance condition by the random change of the quasimomentum. Once an atom loses the resonance condition is stops to follow the ratchet-like directed motion \cite{PhysRevA.94.043620,doi:10.1002/andp.201600335} and hence to follow the designed walk dynamics \cite{PhysRevA.93.023638}. Therefore a steering of the walk by induced SE is not directly possible.}

\begin{figure}[tb]
	\includegraphics[width=0.49\linewidth]{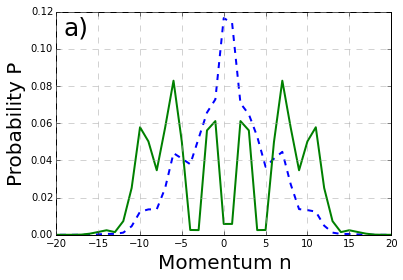}
	\includegraphics[width=0.49\linewidth]{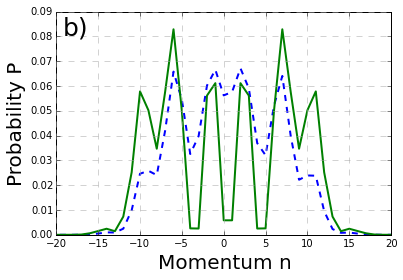}\\
	\includegraphics[width=0.49\linewidth]{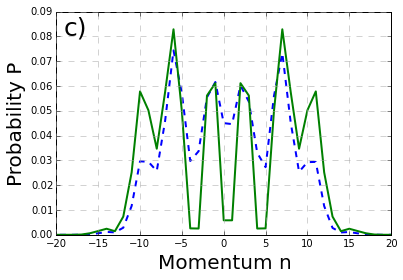}
	\includegraphics[width=0.49\linewidth]{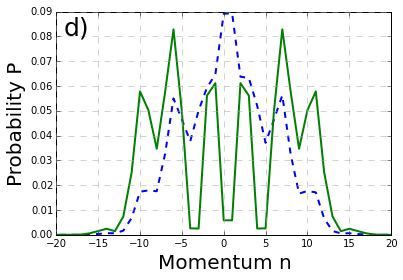}
	\caption{Numerical simulations with $k=1.45$, $T=15$ walk steps (blue dashed lines), for $p_{\rm SE}=0.037, \Delta_\beta=0.025$ (a), $p_{\rm SE}=0.037,  \Delta_\beta=0.01$ (b),  $p_{\rm SE}=0.02, \Delta_\beta=0.01$ (c), and $p_{\rm SE}=0.02, \Delta_\beta=0.02$ (d). The data is averaged over a Gaussian distribution of 1000 quasimomenta, respectively. The green solid line always shows the ideal walk without SE and fixed resonant quasimomentum. Panel (a) and (d) show the trend toward a classical-like distribution centered at zero, which would appear for longer times, just as in panel (d) of Fig. \ref{fig3}.}
	\label{fig5}
\end{figure}

Any experimental implementation based on a Bose-Einstein condensate has a certain width of quasimomentum in the initial state of the experiment \cite{Duffy2, Ryu,  Behinaein_2006, gil2010, gil2010b, PhysRevLett.121.070402, PhysRevA.94.043620, doi:10.1002/andp.201600335, PhysRevE.88.034901,  gil2008, sadgrove2007, gil2012, hoogerland2013}. It turned out that a very good approximation is the modelling of this widths by averaging over a Gaussian distribution of trajectories each of which at fixed initial quasimomentum \cite{Nonl2003, SW2011, gil2012, PhysRevLett.121.070402}. This width measured as full widths at half maximum can be as small as $\Delta_\beta=0.01$ \cite{Ryu}, while
a typical value for the Oklahoma experiment is about $\Delta_\beta=0.02$ \cite{PhysRevLett.121.070402, PhysRevA.94.043620, doi:10.1002/andp.201600335, PhysRevE.88.034901, gil2012}. Since SE mixes the values of quasimomenta, which in the ideal coherent evolution would be preserved, we expect that the effect of SE is now enhanced when starting already from a broad quasimomentum distribution. That this is indeed the case, can be seen in Fig. \ref{fig5}, which presents simulations for a thousand values of quasimomenta and for 15 steps of the walk. The walk becomes classical on average for a width of about $\Delta_\beta \ge 0.025$, largely independent of the SE rates \cite{PhysRevA.93.023638}. 

Altogether, we can state that the limits for observing a coherent quantum walk over a substantial number of kicks are defined by the upper bounds $p_{\rm SE} \le 0.02$ and $\Delta_\beta \le 0.02$. Both ranges are already within experimental reach with $p_{\rm SE} < 0.01$ from \cite{PhysRevE.88.034901} and $\Delta_\beta \approx 0.02$ in the more recent walk experiment \cite{PhysRevLett.121.070402}, for which a similar SE rate probably applies.

%

\section{Conclusions and Outlook}
\label{concl}

In summary, we completed the description of the quantum walks in momentum space with a kicked Bose condensate by setting up a master equation for an important and, to some extend, controllable source of decoherence, namely spontaneous emission.

In the experiment, the microwaves that induce the coin toss and the compensation of the light shift phase and the dynamic phase are not instantaneous but rather work for a finite time, actually the entire period in-between two kicks \cite{PhysRevLett.121.070402}. Since during the free evolution time the internal degrees are not directly coupled that should, however, not be a problem. Much more time consuming simulations on the basis of our \textcolor{black}{new} master equation \textcolor{black}{for the effective three level system}, which is presented here, might check this assumption.

Another approximation was the one of a the 'closed' system composed of three states. Further decay channels outside the three-state system studied here could be included, with explicit loss of atoms during the walk dynamics. From a practical point of view, as explained at the beginning of sec. \ref{sec:3_1} we do expect, however, neither a qualitative nor a significant quantitative change in our results. 

\begin{acknowledgments} 
We thank very much Gil Summy for sharing his insights into the experimental realization and providing Figure 1. Moreover, we are grateful to Scott Parkins for discussions on the effects of spontaneous emission on the atom optics kicked rotor.
\end{acknowledgments}

%
%

%
%
%

\bibliographystyle{apsrev4-1}

%

\end{document}